\documentclass[a4paper]{jpconf}
\pdfoutput=1 
\usepackage{amsmath,amssymb,mathrsfs,graphicx, subfig}
\begin{document}
\title{Multi-Grid Boron-10 detector for large area applications in neutron scattering science}

\author{K.~Andersen$^1$, T.~Bigault$^2$, J.~Birch$^3$, J.-C.~Buffet$^2$, J.~Correa$^2$, P.~van Esch$^2$, B.~Guerard$^2$, R.~Hall-Wilton$^1$, L.~Hultman$^3$, C.~H\"{o}glund$^3$, J.~Jensen$^3$, A.~Khaplanov$^{1,2*}$, O.~Kirstein$^1$, F.~Piscitelli$^2$, C.~Vettier$^{1,4}$}

\address{$^1$European Spallation Source ESS AB, P.O Box 176, SE-221 00 Lund, Sweden}
\address{$^2$Institute Laue Langevin, Rue Jules Horowitz, FR-380 00 Grenoble, France}
\address{$^3$Department of Physics, Chemistry and Biology (IFM), Thin Film Physics Division, Link\"{o}ping University, SE-581 83 Link\"{o}ping, Sweden}
\address{$^4$European Synchrotron Radiation Facility, BP 220, FR-380 43 Grenoble Cedex 9, France}

\ead{*anton.khaplanov@esss.se (corresponding author)}

\begin{abstract}
The present supply of $^3$He can no longer meet the detector demands of the upcoming ESS facility and continued detector upgrades at current neutron sources. Therefore viable alternative technologies are required to support the development of cutting-edge instrumentation for neutron scattering science. In this context, $^{10}$B-based detectors are being developed by collaboration between the ESS, ILL, and Link\"{o}ping University. This paper reports on progress of this technology and the prospects applying it in modern neutron scattering experiments. 
The detector is made-up of multiple rectangular gas counter tubes coated with $B_4C$, enriched in $^{10}$B. An anode wire reads out each tube, thereby giving position of conversion in one of the lateral co-ordinates as well as in depth of the detector. Position resolution in the remaining co-ordinate is obtained by segmenting the cathode tube itself. Boron carbide films have been produced at Link\"{o}ping University and a detector built at ILL. 
The characterization study is presented in this paper, including measurement of efficiency, effects of the fill gas species and pressure, coating thickness variation on efficiency and sensitivity to gamma-rays. 
\end{abstract}

\section{Introduction}

This paper presents our approach to replacing $^3He$ in large-area detectors for neutron scattering instruments. The new technique is based on thin $^{10}B$-containing films. The detector concept builds on the experience with proportional gas detectors, such as are conventionally used in $^3He$ detectors. However, the gaseous neutron converter, is replaced by layers of solid converter~\cite{guerard}. 

In our concept of the multi-grid detector, thin films of boron carbide ($B_4C$) enriched in $^{10}B$ are used as the neutron converter. Since a maximum efficiency of a single such layer is limited to a few percent, many layers have to used~\cite{mcgregor}. $Al$ substrates, that we will refer to as \emph{blades}, support $B_4C$ coatings on each surface. These are placed orthogonally to the direction of incoming neutrons.For some applications it may be advantageous to place the layers at a small angle to the beam\cite{buffet}. We will, however, focus on orthogonal geometry due to the problematics of covering a large area. The intermittent space between the layers is filled with gas and the readout  is provided by an anode wire in the center of each cell. 15 such gas cells (30 thin layers) result in reasonable compromise between the efficiency and the amount of readout electronics and the complexity of the assembly. 

\section{Prototype Multi-Grid $^{10}B$ detector}
\label{sect_proto}

\begin{figure}
  \centering
  \subfloat[]{\label{mounting2}\includegraphics[width=0.4\textwidth]{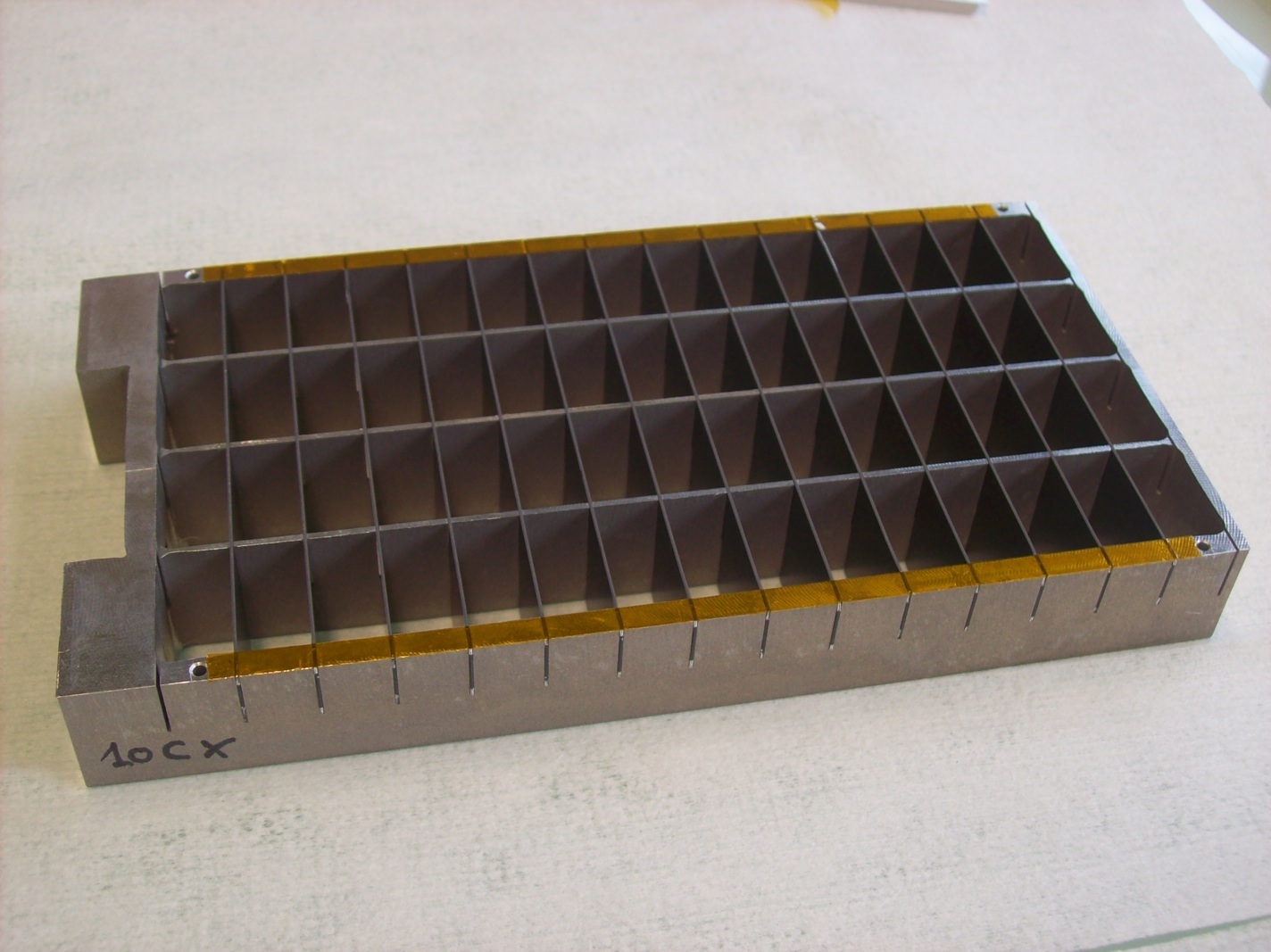}}
  \subfloat[]{\label{mounting1}\includegraphics[width=0.4\textwidth]{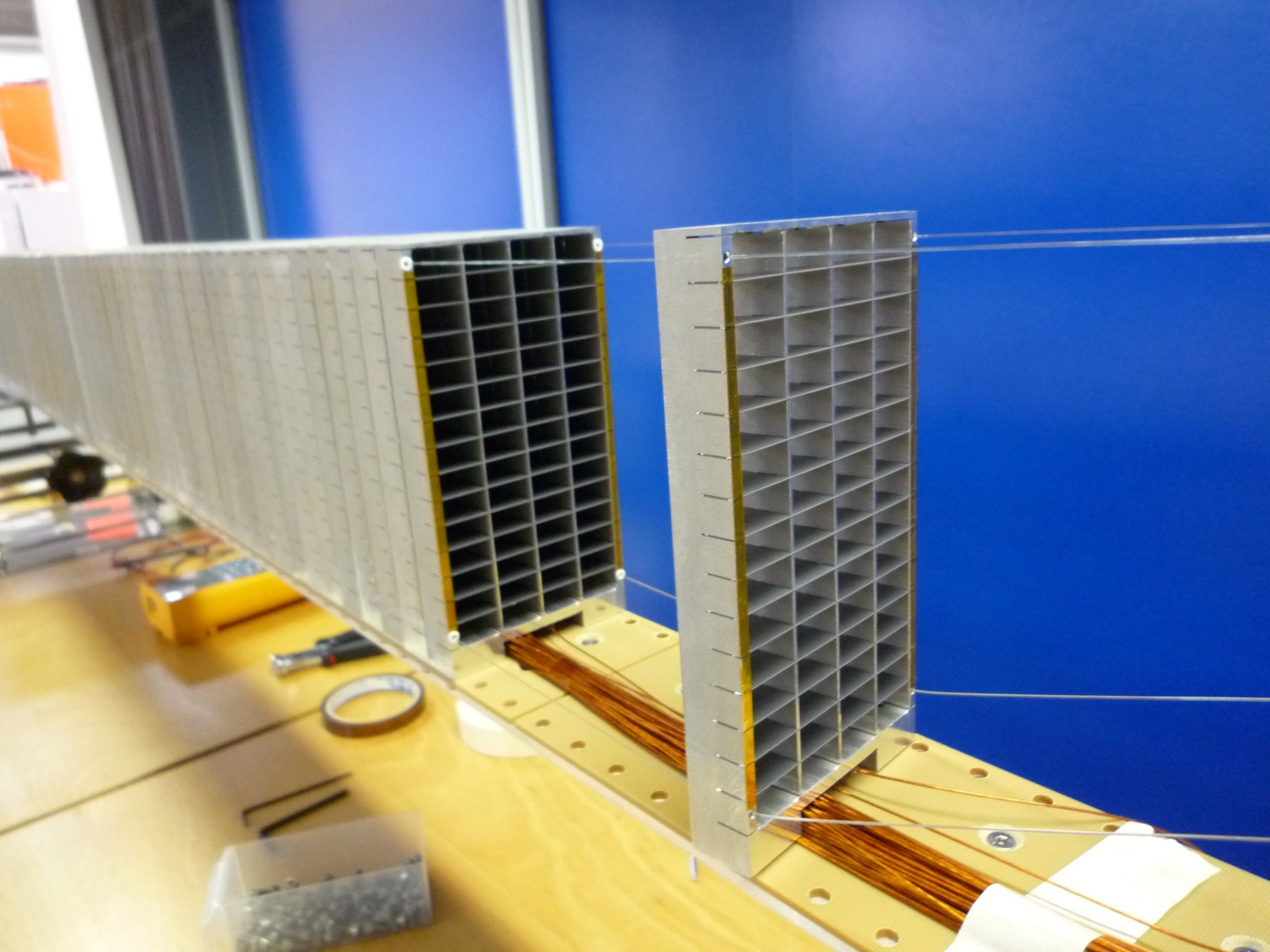}} \\
  \subfloat[]{\label{fit_inbeam}\includegraphics[width=0.8\textwidth]{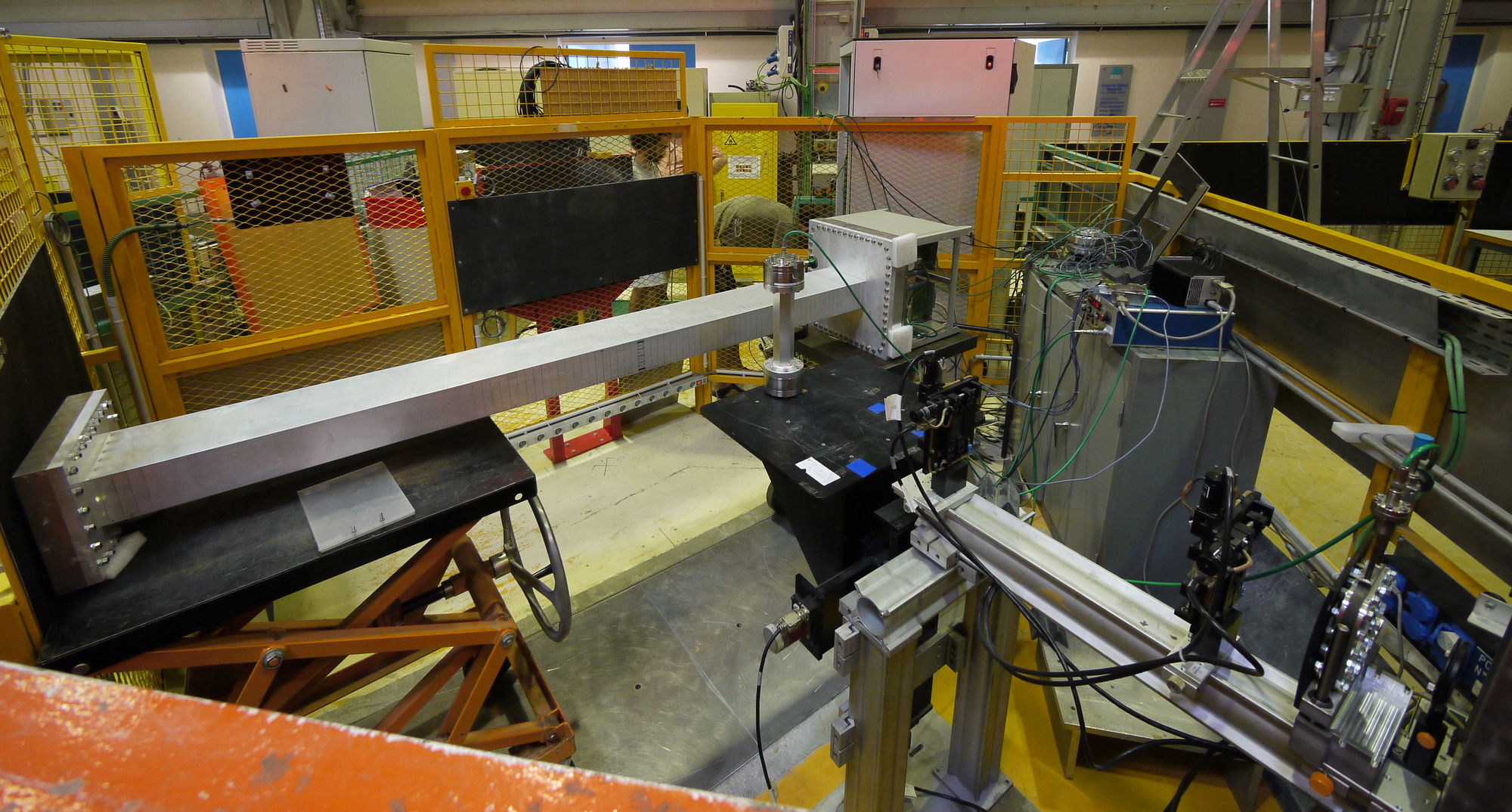}}
  \caption{\textbf{(a)} Grid made of parallel double-coated blades. There are 4 rows of 15 cells which give the depth of the detector is clearly seen. \textbf{(b)} Stack of the different grids one after the other to recreate a tubular shape. \textbf{(c)} The detector at the CT2 test beam line at ILL.}
  \label{fig_proto2}
\end{figure}

The detector is built up of 96 grids. These act as a segmented cathode and support the $^{10}B_4C$ coating. The total active surface is $200 cm \times 8 cm = 0.16 m^{2}$. 14 two-side coated blades ($8 \times 2 cm^{2}$) are used in each grid. A picture of an individual grid and the partially completed detector assembly can be seen in Fig.~\ref{mounting2}. Hence, a total of 28 $B_{4}C$ layers are used for neutron detection. Grids are placed 0.4~mm apart in order to minimize dead area while allowing them to be readout individually. There are 4 rows of 15 anode wires, each traversing the entire length of the frame assembly. Both frame and wire channels were equipped with individual amplifiers. Data can be acquired as singles or frame-wire coincidence. In case of coincidence measurement, a 3-D image is obtained with 4x15x96=5760 voxels.

\section{Producing the $^{10}B_4C$ films}

\begin{figure}[htbp]
\begin{center}
  \subfloat[]{\label{fig_coating}\includegraphics[width=0.35\textwidth]{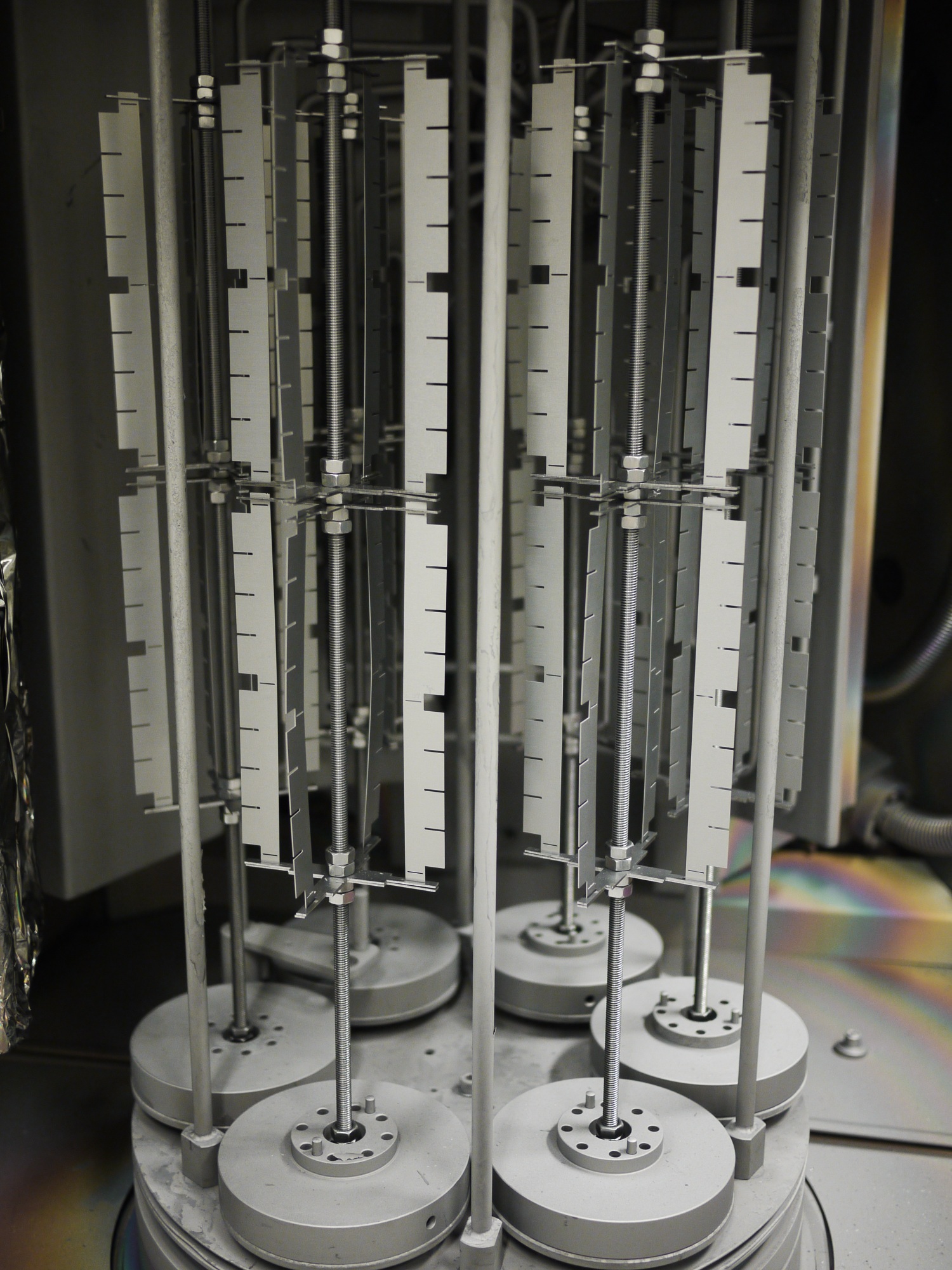}}
  \subfloat[]{\label{fig_gradient}\includegraphics[width=0.64\textwidth]{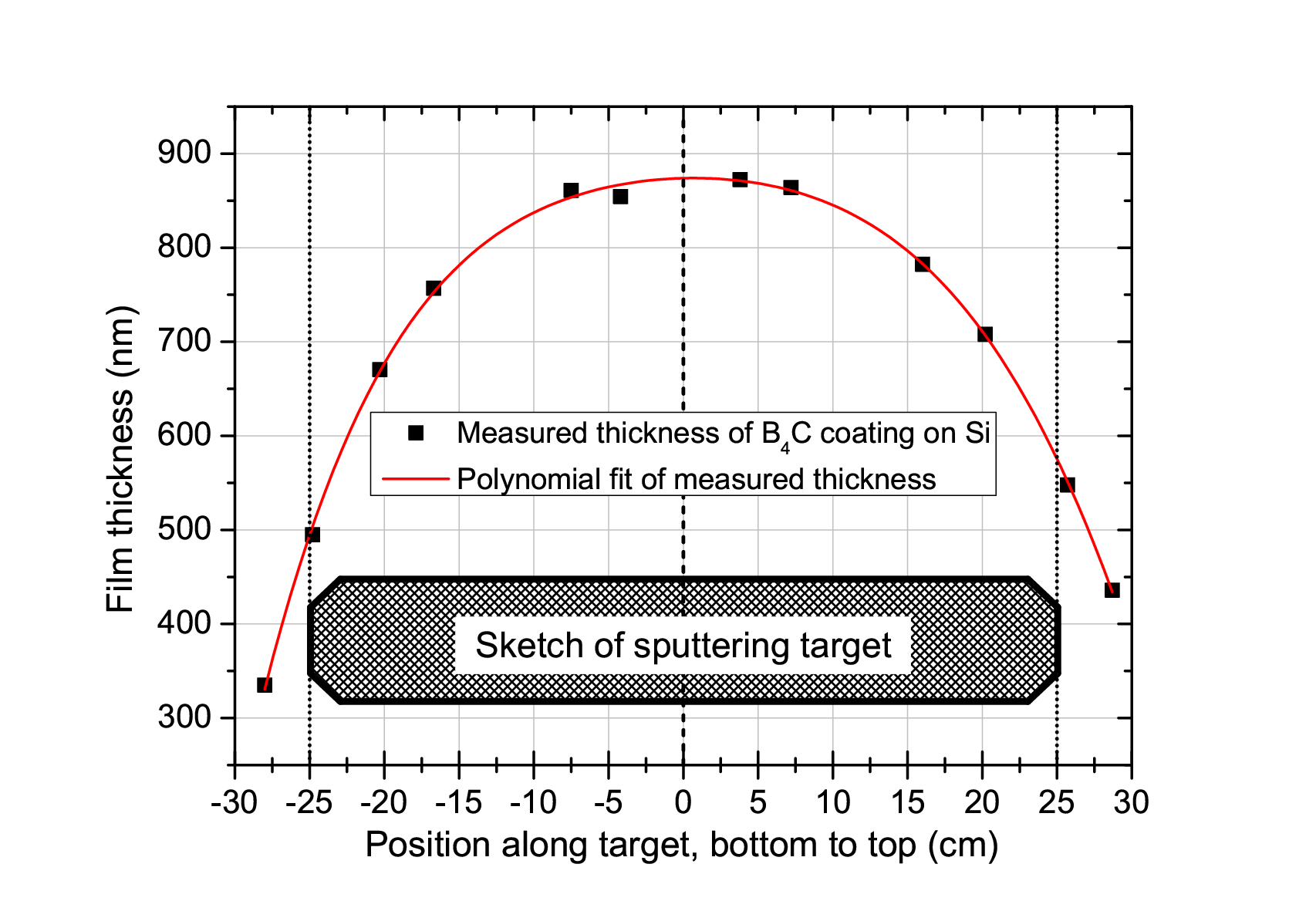}}   
\end{center}
\caption{\textbf{(a)} Newly-coated aluminium blades in the sputtering machine. \textbf{(b)}A typical variation of the coating thickness. The blades where placed between positions -18 and +18 cm of the system.}
\label{fig_gradient}
\end{figure}

Films of $B_4C$ were deposited onto 0.5mm thick aluminium substrate (\emph{blades}) pre-machined to be inserted into the frames~\cite{hoglund}. The deposition process chosen was DC magnetron sputtering. The stuttering chamber is shown in fig.~\ref{fig_coating}. A 97\% enriched sputtering target was used. The isotopic composition of final coatings was measured using time-of-flight elastic recoil detection analysis (ToF-ERDA) method; coating density -- by x-ray reflectivity analysis; and coating thickness -- by examining reference samples coated onto silicon in an scanning electron microscope (SEM). Our optimization of the coating thickness uses the measured parameters of the coating as well as a 100~keV detection threshold. This optimization follows the principle presented in detail in~\cite{mcgregor}. A neutron wavelength of 2.5~\AA~ was chosen for optimization as the wavelength that is available at the CT1 and CT2 test beams at ILL. The resulting optimal thickness is $1.0 \mu m$ and detection efficiency is 58\% for 28 conversion layers. This is reduced to 53\% when the scattering in the $Al$ vessel and substrates is included. 

The coatings produced show a total thickness variation of about 11\%. This is illustrated in Fig.~\ref{fig_gradient}. The deposition chamber is 50~cm tall. The 8~cm long blades are placed vertically in four levels occupying the middle section of the chamber. A considerable gradient in coating thickness in top- and bottom-most blades is expected. The maximum variation due to this effect does not exceed 8\% over the sensitive area of a blade. While for blades in the center of the deposition chamber this value is below 2\%. This was furthermore verified by neutron absorption measurements. 

The coating thickness variation is systematic and having marked the location of each blade during the coating process, it was then possible to compensate any non-uniformity in efficiency by assembling the frames in a way that the the two coatings contributing to each detector cell compensate each other's thickness gradient. A fraction of the frames were deliberately assembled without this compensation in order to quantify this effect.

\section{Detecting gas and pressure}

\begin{figure}[htbp]
\begin{center}
\resizebox{0.6\textwidth}{!}{\includegraphics{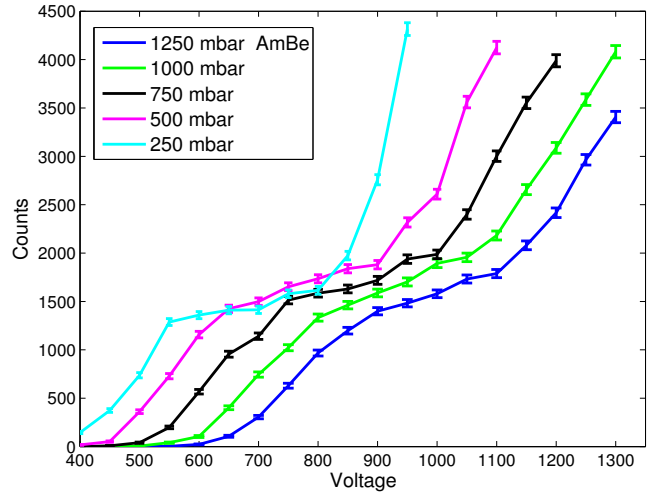}}
\end{center}
\caption{Detector plateau for different pressures of $Ar-CO_2$ measured with $AmBe$ neutron source. Note that the source also emits a high flux of $\gamma$-rays, predominantly at low energies. These can be seen as the sharp increase in the count rate at a certain bias voltage. The electronic threshold is constant in all measurements.}
\label{fig_plateau}
\end{figure}

The detector has been operated with either of two gasses: $CF_4$ or $Ar/CO_2$ mixture (90\%-10\%). The performance is comparable, however the $Ar/CO_2$ allows operation at a lower bias voltage. In contrast to $^3He$ detectors, where the pressure is often in the order of 5~bar, the $^{10}B$ detector can be operated at atmospheric pressure or lower. This makes it particularly well-suited for instruments where the entire detector assembly is placed in vacuum. Fig.~\ref{fig_plateau} shows the behavior of the prototype at varying pressure.

\section{Efficiency measurement}

\begin{figure}[htbp]
\begin{center}
\resizebox{0.6\textwidth}{!}{\includegraphics{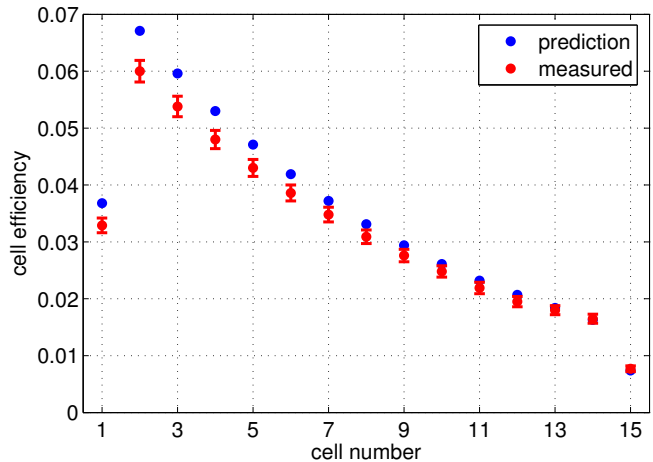}}
\end{center}
\caption{Partial efficiency of a row of 15 cells of the prototype for a incident beam of 2.5~\AA neutrons traversing the cells from left to right. Note the approximately halved efficiency in the first and last cell. This is due to only one layer of $^{10}B_4C$ contributing to the detections in these.}
\label{fig_profile}
\end{figure}

Neutron efficiency of the detector was measured at the monochromatic 2.5~\AA~ beam at CT2 at ILL. The beam was collimated to a $2 \times 2 mm$ spot. The stability of the neutron flux was verified with a beam monitor. In these fixed conditions a rate measurement was performed for the $^{10}B$ prototype and compared against an absolutely calibrated $^3He$ detector. The detector was filled with 1~bar of $CF_4$, the bias voltage was set to 1700V. Counts were acquired for coincident signal between the anodes and cathodes of the detector. In this way voxels where the detection occurred are determined. Random electronic noise is also efficiently rejected. The efficiency was measured to be 47.9 +-0.3\%(stat.). As the beam was narrowly collimated we are confident that vast majority of conversions occurred in only one row of voxels. Fig.~\ref{fig_profile} shows the partial contribution of the 15 voxel to the detector efficiency. A prediction based on calculation is shown for comparison. 

The exact definitions of \emph{counts} is somewhat complicated by a considerable crosstalk between the cathode frames. Crosstalk between wires is negligible. For sufficiently large signals the induced signals in the neighbouring channels exceed the threshold and are sometimes captured by the data acquisition system instead of the signal in the real target frame. These situation are not trivially quantified since another reason to find signals in the neighbouring frames is the scattering of the neutron beam in the Al substrate and air. A somewhat higher efficiency could be obtained if the acquisition system was able to sample more than one frame for each event and choose the one with the highest signal. A further addition to the systematic error is the scattering effects and signal pile-up in the $3He$ reference detector. The overall systematic error is $\{+2.6, -1.1\}$.

\section{Efficiency uniformity}

In order to investigate the effect of the variation of coating thickness among the blades used a scan of the entire detector was performed. An $AmBe$ neutron source was translated along the detector at a uniform rate over a period of several days, resulting in a uniform time-average flux on the detector. 

As described in section~\ref{sect_proto}, detector frames were assembled using 14 blades. Based on information in fig.~\ref{fig_gradient}, several different types of frames were assembled based on the maximum gradient expected in the blades used. In most frames, blade orientation was alternated, so that for each detector cell, the thickness (and, therefore, efficiency) varies in opposite directions in the front and rear coating. This largely cancels the overall effect on efficiency across that frame. In a number of the frames, however, this compensation was deliberately not implemented. Also the blades with the greater gradient were used in these frames. Therefore we expect up to 8\% variation in total amount of $^{10}B$ from one side of these frames to the other. 

\begin{figure}[htbp]
\resizebox{0.98\textwidth}{!}{\includegraphics{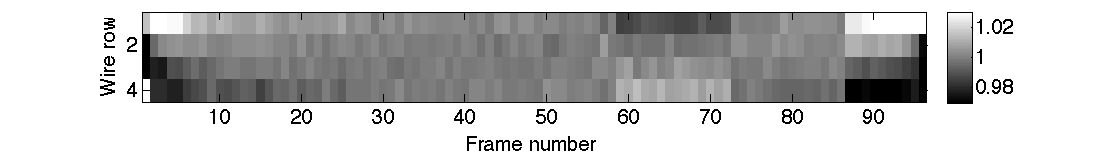}}
\caption{Source scan of the detector. Intensity in each pixel is the sum of the contributing 15 cells. The two regions where orientation of the blades was not alternated are at frames 59 -- 72 and 86 -- 96 The first is clearly visible and corresponds to a $\pm 1\%$ efficiency variation. The second is partially obscured by an edge effect which is also seen in the left end of the detector. This is due to shape of the shielding around the source and the fact that the source did not uniformly irradiate several edge frames.}
\label{fig_uni}
\end{figure}

As shown in fig.~\ref{fig_uni}, a deviation from a uniform response was found in frames where the blade orientation was not alternated. This variation is on average approximately 2\% -- in good agreement with the expectation based on the coating process and efficiency calculation described above. 

In should be noted that the 2\% efficiency variation derived above is the highest possible over the sensitive area of the detector due to coating thickness (effect of the electronics is excluded from this analysis). The efficiency in the majority of frames where blade orientation was alternated is on the level of statistical variation. For the non-compensated frames, however, one side of the frame has the thinnest coating in the detector and the other -- the thickest. With the use of the simple idea of alternating blade direction we can therefore achieve a uniformity of better than 1\% for the entire active surface of the detector.


\section{Gamma sensitivity}

\begin{figure}[htbp]
\begin{center}
\resizebox{0.6\textwidth}{!}{\includegraphics{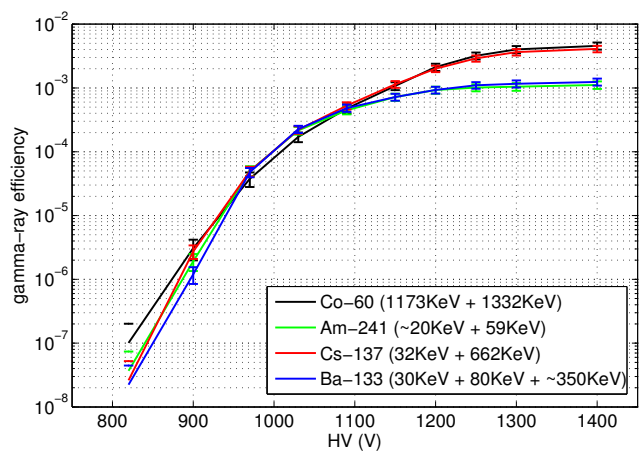}}
\end{center}
\caption{Measurement of gamma response for a range of sources and bias voltages. Note that the value at the lowest voltage used is an upper limit -- here, no signal could be detected over statistical variation.}
\label{fig_gamma}
\end{figure}

The discrimination of gamma contamination of the neutron response is not as straight-forward for the multi-grid detector as for a conventional $^3He$ tube. In order to quantify the trade-off between neutron efficiency and gamma rejection measurements have been performed on the prototype with gamma sources in the absence of neutron sources. The results are presented here as gamma detection efficiency. This can then be easily compared to neutron efficiency. Measurements were performed for 4 different gamma-emitters at several different voltages and are summarized in fig.~\ref{fig_gamma}. The gamma efficiencies take into account the activities of the sources, relative gamma intensities and absorption of the x-rays and lower-energy gamma-rays in the aluminium wall of the pressure vessel.

The $\gamma$ sources were chosen to have a wide coverage of energies. The response to the significantly varying photon energy is deceptively similar in terms of the measured pulse heights, however. This can be understood by considering the geometry of the detector. Photo or Compton electrons produced in either walls of the grids or the gas, will usually travel at most 2~cm before encountering a wall. The energy deposited in the gas and readout by a single grid is therefore limited. Based on our measurements this limit is below 100~keV. 

For the lowest voltage used, only an upper limit of gamma efficiency can be obtained since no signal could be detected over the background. This setting corresponds to an approximately 41\% efficiency at 2.5~\AA~ (a reduction of about 7\% compared to our in-beam measurement), however yields better than $10^{-6}$ gamma rejection. 

The measurements of both gamma and neutron efficiencies were performed at a pressure of 1~bar. A lower pressure would improve the neutron to gamma ratio since photo and Compton electrons will deposit less energy in the gas before stopping in a wall. This is also true of the neutron conversion products, however their energy deposition will saturate at considerably higher than that of the electrons. Indeed, operation at a lower than atmospheric pressure is likely to be relevant since in many experiments detectors need to be placed in a vacuum vessel.

\section{Conclusions}

Throughout the design, assembly and characterization of our $^{10}B$ prototype, it has been shown that this detector technology is a viable replacement of the no longer available $^3He$. It has been shown that detectors of large dimensions can be built for neutron scattering experiments that show an intrinsic efficiency of 50\% (this will vary with wavelength), better that 1\% uniformity, provide the necessary position sensitivity  and are well-suited for operation in vacuum. The sensitivity to $\gamma$-rays is greater than in the case of conventional $^3He$ detectors. A good knowledge of the background radiation in the future instruments will therefore be important for defining the optimal operational settings of the $^{10}B$ detector.

\end{document}